\documentclass[aps,prl,twoside,twocolumn,floatfix,amsmath,showpacs,amssymb,nofootinbib]{revtex4}

 \usepackage[latin1]{inputenc}
 \usepackage{amsmath}
 \usepackage{bm}
 \usepackage{subfigure}
 \usepackage{placeins}
 \usepackage{floatflt} \usepackage{amsmath}
 \usepackage{amssymb}
 \usepackage{wrapfig}
 \usepackage[]{tocbibind}
 \usepackage[all]{xy}

 \usepackage{ifthen}
 \usepackage{times}
 
\makeatletter
\newcommand\figcaption{\def\@captype{figure}\caption}
\newcommand\tabcaption{\def\@captype{table}\caption}
\makeatother

\usepackage[pdftex]{graphicx}     %USE WHEN DOCUMENT IS PDFLATEX'ED!
\newcommand{\AuAu}    {Au\,+\,Au collisions at 1.23\agev}   % before
\newcommand{\agev}    {\mbox{$A$~GeV}}               %PRL notation

\newcommand{\mev}     {\mbox{MeV}}
\newcommand{\mevc}    {\mbox{MeV$/c$}}

\newcommand{\rb}[1]   {\mbox{\textrm{\scriptsize #1}}}

\newcommand{\pimin}   {\ensuremath{\pi^{-}}}
\newcommand{\piplus}  {\ensuremath{\pi^{+}}}

\newcommand{\rinv}    {\ensuremath{R_{\rb{inv}}}}

\newcommand{\qinv}    {\ensuremath{q_{\rb{inv}}}}
\newcommand{\qout}    {\ensuremath{q_{\rb{out}}}}
\newcommand{\qside}   {\ensuremath{q_{\rb{side}}}}
\newcommand{\qlong}   {\ensuremath{q_{\rb{long}}}}

\newcommand{\rout}    {\ensuremath{R_{\rb{out}}}}
\newcommand{\rside}   {\ensuremath{R_{\rb{side}}}}
\newcommand{\rlong}   {\ensuremath{R_{\rb{long}}}}
\newcommand{\routlong}   {\ensuremath{R_{\rb{out\,long}}}}
\begin{document}
\title{Identical pion intensity interferometry in central Au\,+\,Au collisions at 
1.23\boldmath{$A$}~GeV}
\author{
 J.~Adamczewski-Musch$^{4}$, O.~Arnold$^{10,9}$, C.~Behnke$^{8}$, A.~Belounnas$^{16}$,
A.~Belyaev$^{7}$, J.C.~Berger-Chen$^{10,9}$, J.~Biernat$^{3}$, A.~Blanco$^{2}$, C.~~Blume$^{8}$,
M.~B\"{o}hmer$^{10}$
% ...
\\ \vskip 10bp (HADES collaboration)}

\affiliation{
\mbox{ } \\
\mbox{$^{1}$Institute of Physics, Slovak Academy of Sciences, 84228~Bratislava, Slovakia}\\
\mbox{$^{2}$LIP-Laborat\'{o}rio de Instrumenta\c{c}\~{a}o e F\'{\i}sica Experimental de Part\'{\i}culas , 3004-516~Coimbra, Portugal}\\
\mbox{$^{3}$Smoluchowski Institute of Physics, Jagiellonian University of Cracow, 30-059~Krak\'{o}w, Poland}\\
\mbox{$^{4}$GSI Helmholtzzentrum f\"{u}r Schwerionenforschung GmbH, 64291~Darmstadt, Germany}\\
\mbox{$^{5}$Technische Universit\"{a}t Darmstadt, 64289~Darmstadt, Germany}\\
\mbox{$^{6}$Institut f\"{u}r Strahlenphysik, Helmholtz-Zentrum Dresden-Rossendorf, 01314~Dresden, Germany}\\
\mbox{$^{7}$Joint Institute of Nuclear Research, 141980~Dubna, Russia}\\
\mbox{$^{8}$Institut f\"{u}r Kernphysik, Goethe-Universit\"{a}t, 60438 ~Frankfurt, Germany}\\
\mbox{$^{9}$Excellence Cluster 'Origin and Structure of the Universe' , 85748~Garching, Germany}\\
\mbox{$^{10}$Physik Department E62, Technische Universit\"{a}t M\"{u}nchen, 85748~Garching, Germany}\\
\mbox{$^{11}$II.Physikalisches Institut, Justus Liebig Universit\"{a}t Giessen, 35392~Giessen, Germany}\\
\mbox{$^{12}$Institute for Nuclear Research, Russian Academy of Science, 117312~Moscow, Russia}\\
\mbox{$^{13}$Institute of Theoretical and Experimental Physics, 117218~Moscow, Russia}\\
\mbox{$^{14}$National Research Nuclear University MEPhI (Moscow Engineering Physics Institute), 115409~Moscow, Russia}\\
\mbox{$^{15}$Department of Physics, University of Cyprus, 1678~Nicosia, Cyprus}\\
\mbox{$^{16}$Institut de Physique Nucl\'{e}aire, CNRS-IN2P3, Univ. Paris-Sud, Universit\'{e} Paris-Saclay, F-91406~Orsay Cedex, France}\\
\mbox{$^{17}$Nuclear Physics Institute, The Czech Academy of Sciences, 25068~Rez, Czech Republic}\\
\mbox{$^{18}$LabCAF. F. F\'{\i}sica, Univ. de Santiago de Compostela, 15706~Santiago de Compostela, Spain}\\
\mbox{$^{19}$Uniwersytet Warszawski - Instytut Fizyki Do\'{s}wiadczalnej, 02-093 Warszawa, Poland}\\
\\
\mbox{$^{a}$ also at Coimbra Polytechnic - ISEC, ~Coimbra, Portugal}\\
\mbox{$^{b}$ also at ExtreMe Matter Institute EMMI, 64291~Darmstadt, Germany}\\
\mbox{$^{c}$ also at Technische Universit\"{a}t Dresden, 01062~Dresden, Germany}\\
\mbox{$^{d}$ also at Frederick University, 1036~Nicosia, Cyprus}\\
\mbox{$^{e}$ also at Dipartimento di Fisica and INFN, Universit\`{a} di Torino, 10125~Torino, Italy}
\\ $^{\dag}$ Deceased.
}
 
%\date{Received: 23 January 2018 / Revised: 10 April 2018}
\date{Received \today}
\begin{abstract}
\vskip 15bp
We investigate identical pion HBT intensity interferometry in central \AuAu. 
High-statistics $\pi^-\pi^-$ and $\pi^+\pi^+$ data are measured with HADES at SIS18/GSI. 
The radius parameters, derived from the correlation function depending on relative momenta 
in the longitudinally comoving system and parametrized as three-dimensional Gaussian 
distribution, are studied as function of transverse momentum. 
A substantial charge-sign difference of the source radii is found, particularly  
pronounced at low transverse momentum. The extracted 
source parameters agree well with a smooth extrapolation of the center-of-mass energy 
dependence established at higher energies, 
extending the corresponding excitation functions down towards a very low energy. 
%Our data would thus rather disfavour any strong energy dependence of the radius parameters 
%in the low energy region.  
\end{abstract} 


\maketitle

Two-particle intensity interferometry of hadrons is widely used to study
the spatio-temporal size, shape and evolution of their sources created in
heavy-ion collisions or other reactions involving hadrons 
(for a review see Ref.\,\cite{Lisa05}). 
The technique, pioneered by Hanbury Brown and Twiss \cite{HBT56} to measure angular 
radii of stars, later on named HBT interferometry, is based on the quantum-statistical 
interference of identical particles. 
Goldhaber et al. \cite{GGLP60} first applied intensity interferometry to hadrons.   
In heavy-ion collisions,  
the intensity interferometry does not allow to measure directly the reaction volume,
as the emission source, changing in shape and size in the course of the collision, 
is affected by density and temperature gradients and dynamically generated
space-momentum correlations ({\it e.g.} radial expansion after the compression phase 
or resonance decays).
Thus, intensity interferometry generally does not yield the proper source size, but rather
an effective ``length of homogeneity'' \cite{Lisa05}. It measures source regions in which 
particle pairs are close in momentum, so that they are correlated as a consequence of 
their quantum statistics or due to their two-body interaction.
In general, the sign and strength of the correlation is affected by (i) the
strong interaction, (ii) the Coulomb interaction if charged particles
are involved, and (iii) the quantum statistics in the case of identical
particles (Fermi-Dirac suppression for fermions, Bose-Einstein enhancement for
bosons). In the case of $\pi\pi$ correlations, the mutual strong interaction was found to be 
minor \cite{Bowler88} compared to the effects (ii) and (iii). 

Pion freeze-out dynamics may be relevant to 
ongoing searches for the QCD critical point in the $T-\mu_B$ plane, where $T$ and $\mu_B$ are 
the temperature and the baryon-chemical potential. Systems with $\mu_B$ above the critical 
point are expected to undergo a first-order phase transition which might be visible in a 
non-monotonic behavior of various source parameters. However, it is also conceivable that the 
initial temperatures of the system, which can be reached in heavy-ion collisions at high $\mu_B$, 
are not high enough to create a deconfined partonic state.  In this scenario a first order phase 
boundary cannot be reached experimentally. A recently published excitation function of HBT source 
radii \cite{STAR2015} from the domain of the Relativistic Heavy Ion Collider (RHIC) 
down to lower collision energies indicates such a non-monotonic energy dependence   
around center-of-mass energies of $\sqrt{s_\mathrm{NN}} < 10$~GeV. Even though a part of 
this behavior can be related to the strong impact of different pair transverse 
momentum intervals involved in the source parameter compilation of Ref.\,\cite{STAR2015}, 
to a certain extent the deviation of the data points from a monotonic trend 
remains at low energies. Here, new precision data, especially at low collision energies 
of $\sqrt{s_\mathrm{NN}} < 5$~GeV, can contribute to the clarification of this exciting 
observation before definite conclusions on a change in physics can be drawn.         

It is worth emphasizing that only preliminary data \cite{hbt_fopi_1995} of 
identical-pion HBT data exist for a large symmetric collision system (like Au\,+\,Au or 
Pb\,+\,Pb) at a beam kinetic energy of about $1\agev$ 
(fixed target, $\sqrt{s_\mathrm{NN}} = 2.3$~GeV)\footnote{Throughout this publication 
$\agev$ refers to the mean kinetic beam energy.}.  
For the somewhat smaller system La\,+\,La, studied at $1.2A$~GeV with the HISS spectrometer 
at the Lawrence Berkeley Laboratory (LBL) Bevalac, pion correlation data were reported by 
Christie et al. \cite{Christie93,Christie92}. 
An oblate shape of the pion source and a correlation of the source 
size with the system size were found. Also, pion intensity interferometry for small systems 
(Ar\,+\,KCl, Ne\,+\,NaF) was studied at 1.8$A$~GeV at the LBL Bevalac using the Janus 
spectrometer by Zajc et al. \cite{Zajc84}. Both groups made first  
attempts to correct the influence of the pion-nuclear Coulomb interaction on the pion momenta. 
The effect on the source radii, however, were found negligible for their experiments. 

In this letter we report on the first investigation of $\pi^-\pi^-$ and $\pi^+\pi^+$  
correlations at low relative momenta in \AuAu, continuing our previous femtoscopic studies of 
smaller collisions systems \cite{hades_pLambda_ArKCl,hades_hbt_ArKCl,hades_pLambda_pNb}. 
The experiment was performed with the 
{\textbf H}igh {\textbf A}cceptance {\textbf D}i-{\textbf E}lectron {\textbf S}pectrometer 
(HADES) at the Schwerionensynchrotron SIS18 at GSI, Darmstadt. HADES \cite{Agakishiev:2009am}, 
although primarily optimized to measure di-electrons \cite{HADES-PRL07}, offers also 
excellent hadron identification capabilities 
\cite{hades_kpm_phi_arkcl,hades_xi_arkcl,hades_K0_ArKCl,hades_Lambda_ArKCl}.
HADES is a charged particle detector consisting of a six-coil toroidal magnet centered 
around the beam axis and six identical detection sections located between the coils and
covering polar angles between $18^{\circ}$ and $85^{\circ}$.  Each
sector is equipped with a Ring-Imaging Cherenkov (RICH) detector
followed by four layers of Mini-Drift Chambers (MDCs), two in
front of and two behind the magnetic field, as well as a scintillator
Time-Of-Flight detector (TOF) ($45^{\circ}$~--~$85^{\circ}$) and
Resistive Plate Chambers (RPC) ($18^{\circ}$~--~$45^{\circ}$). Both timing detectors, TOF and 
RPC, allow for good particle identification, i.e. proton-pion separation. (Due to their low 
yield, kaons hardly affect the pion selection at SIS energies.)    
TOF, RPC, and Pre-Shower detectors (behind RPC, for e$^\pm$ identification) 
were combined into a Multiplicity and Electron Trigger Array (META). 
Several triggers are implemented.  The
minimum bias trigger is defined by a signal in a diamond START detector
in front of the 15-fold segmented gold target.  
In addition, online Physics Triggers (PT) are used, which
are based on hardware thresholds on the TOF signals, proportional to
the event multiplicity, corresponding to at least 20 (PT3)
hits in the TOF.  
About 2.1 billion PT3 triggered 
Au\,+\,Au collisions corresponding to the 40\,\% most central events 
are taken into account for the correlation analysis. 
The centrality determination is based on the summed number of hits 
detected by the TOF and the RPC detectors. The measured events are 
divided in centrality classes corresponding to successive $10\,\%$ regions  
of the total cross section \cite{hades_centrality:2018am}. 
Here, we report only on results of the $0-10\,\%$ class; 
the entire centrality dependence of pion source parameters analysed as function 
of azimuthal angle w.r.t. the reaction plane will be part of an extended 
forthcoming paper, while yields and phase-space distributions of charged pions 
are to be presented in a separate report. 

Generally, the two-particle correlation function is defined as the ratio of the 
probability $P_2({\bm p}_1,{\bm p}_2)$ to measure simultaneously two particles with 
momenta ${\bm p}_1$ and ${\bm p}_2$ and the product of the corresponding single-particle 
probabilities $P_1({\bm p}_1)$ and $P_1({\bm p}_2)$ \cite{Lisa05}, 
\begin{equation}
C({\bm p}_1, {\bm p}_2) = \frac{P_2({\bm p}_1,{\bm p}_2)}{P_1({\bm p}_1)  P_1({\bm p}_2)}.
\label{def_theo_corr_fct}
\end{equation}
Experimentally this correlation is formed as a function of the momentum difference between the 
two particles of a given pair and quantified by taking the ratio of the yields of 'true' pairs 
($Y_\mathrm{true}$) and uncorrelated pairs ($Y_\mathrm{mix}$). $Y_\mathrm{true}$ is 
constructed from all particle pairs in the selected phase space interval from the same event.
$Y_\mathrm{mix}$ is generated by event mixing, where particle 1 and particle 2 are taken 
from different events. Care was taken to mix particles from similar event classes in terms of 
multiplicity, vertex position and reaction plane angle. 
The events are allowed to differ by not more than 10 units in the number of 
the RPC\,+\,TOF hit multiplicity of $\ge182$ (i.e. corresponding to the 
uncertainty of the centrality class $0-10\,\%$ \cite{hades_centrality:2018am}), 1.2\,mm in 
the $z$-vertex coordinate (amounting to less than one third of the spacing between target 
segments), and 30 degrees in azimuthal angle (to be compared to the event plane resolution 
of $\langle \cos{\Phi}\rangle=0.612$), respectively. 

The momentum difference is decomposed into three orthogonal components as suggested by 
Podgoretsky~\cite{Podgoretsky83}, Pratt~\cite{Pratt86} and Bertsch~\cite{Bertsch89}.  
The three-dimensional correlation functions are projections of 
equation (\ref{def_theo_corr_fct}) into the (out, side, long)-coordinate system, 
where `out' means along the pair transverse momentum,  
${\bm k}_\mathrm{t}=({\bm p}_\mathrm{t,\,1}+ {\bm p}_\mathrm{t,\,2})/2$, `long' is 
parallel to the beam direction z, and `side' is oriented perpendicular to the other 
directions. The particles forming a pair are boosted into the longitudinally comoving  
system (LCMS), where the z-components of the momenta cancel each other, 
$p_\mathrm{z_1}+p_\mathrm{z_2}=0$. Note that in other publications 
also the pair comoving system (${\bm p}_1+{\bm p}_2=0$) is frequently used.  
The LCMS choice allows for an adequate comparison with correlation data taken at very 
different, usually much higher, collision energies, where the distribution of the rapidity, 
$y=\tanh^{-1}{(\beta_\mathrm{z})}$, of produced particles is found to be not as narrow as in 
the present case but largely elongated. (Here, $\beta_\mathrm{z}=p_\mathrm{z}/E$, 
$E=\sqrt{p^2+m_0^2}$ and $m_0$ are the longitudinal velocity, 
the total energy and the rest mass of the particle, respectively. 
We use units with $\hbar=c^2=1$.)
Hence, the experimental correlation function is given by  
\begin{equation}
C(\qout,\qside,\qlong) = {\cal N} \,\frac{Y_{\mathrm{true}}(\qout,\qside,\qlong)}{Y_{\mathrm{mix}}(\qout,\qside,\qlong)},  
\label{def_exp_corr_fct}
\end{equation}
where $q_i=(p_{1,\,i}-p_{2,\,i})/2$ ($i$\,=\,'out',\,'side',\,'long') are the relative 
momentum components, and ${\cal N}$ is a normalization factor which is fixed by the 
requirement $C \rightarrow 1$ at large relative momenta, where the
correlation function is expected to flatten out at unity. 
Note that, as in our previous intensity interferometry analyses 
\cite{hades_pLambda_ArKCl,hades_hbt_ArKCl,hades_pLambda_pNb}, 
we use the above low-energy convention of $q$ 
which is common also in studies of proton-proton correlations, 
in contrast to the high-energy convention of $\pi\pi$ correlations, $Q=2q$. 
The statistical errors of 
equation (\ref{def_exp_corr_fct}) are dominated by those of the true yield, 
since the mixed yield is generated with much higher statistics.

Two-track reconstruction defects (e.g. track splitting and merging effects) that are 
particularly important to HBT analyses were corrected by appropriate selection conditions on 
the META-hit and MDC-layer levels, i.e. by discarding pairs which hit the same META cell, 
and by excluding for particle\,2 three successive wires symmetrically around the MDC wire 
fired by particle\,1. This method was tested with simulations carrying neither 
quantum-statistical nor Coulomb effects, 
based on UrQMD \cite{UrQMD}, Geant\,\cite{GEANT} and a detailed 
description of the detector response, to firmly exclude any close-track effect. 
Also broader exclusion windows have been tested, but no significant 
improvement was found. These simulations also showed that there are no significant long-range 
correlations, usually attributed either to energy-momentum conservation in correlation analyses 
of small systems or to minijet-like phenomena at high energies.  

The data are divided into seven $k_\mathrm{t}$ bins from 50 to 400~$\mevc$. 
The three-dimensional experimental correlation function is then fitted with the function
\begin{align}
& C_\mathrm{fit}(\qout,\,\qside,\,\qlong) = \nonumber \\ 
& N \big[(1-\lambda)  + \lambda\, K_\mathrm{C}(\hat q,\rinv) \,C_\mathrm{qs}(\qout,\,\qside,\,\qlong)\big], 
\label{pipi_fit_fct_3dim}
\end{align} 
where 
\begin{align}
& C_\mathrm{qs}(\qout,\,\qside,\,\qlong)= 1\,+ \nonumber \\ 
& \exp{(-(2\qout\rout)^2-(2\qside\rside)^2-(2\qlong\rlong)^2)} \hfill
\label{pipi_fit_be_3dim}
\end{align} 
represents the quantum-statistical part of the correlation function. 
The parameters $N$ and $\lambda$ in Eq.\,(\ref{pipi_fit_fct_3dim})
are a normalization constant and the fraction of 
correlated pairs, respectively, and $\hat q= \qinv(\qout,\,\qside,\,\qlong,\,k_\mathrm{t})$
is the average value of the invariant momentum difference, 
$\qinv =\frac{1}{2}\sqrt{({\bm p}_1 - {\bm p}_2)^2-(E_1-E_2)^2}$, 
for given intervals of the relative momentum components and $k_\mathrm{t}$. The range of 
the one- and three-dimensional fits extends in $\qinv$ from $6\,\mevc$ to $80\,\mevc$. 
Log-likelihood minimization \cite{Ahle02} was used in all fits to the correlation functions. 
The influence of the mutual Coulomb interaction in Eq.\,(\ref{pipi_fit_fct_3dim}) is 
separated from the Bose-Einstein part by including in the fits the commonly used Coulomb 
correction by Sinyukov et al.\,\cite{Sinyukov98}. 
The Coulomb factor $K_\mathrm{C}$ results from the integration of the 
two-pion Coulomb wave function squared over a spherical Gaussian source of fixed radius.   
This radius is iteratively approximated by the result of the corresponding fit to the 
correlation function. 
In Eq.\,(\ref{pipi_fit_fct_3dim}), the non-diagonal elements comprising 
the combinations 'out'-'side' and 'side'-'long' vanish for symmetry reasons \cite{UHeinz02}  
when azimuthally and rapidity integrated  
correlations functions are studied \cite{e895_2000,HBT_in_UrQMD}, 
as it is done in the present investigation. The 'out'-'long' component, however, 
can have a finite value depending on the degree of symmetry 
of the detector-accepted rapidity distribution w.r.t. midrapidity ($y_\mathrm{cm}=0.74$). 
We studied this effect by including in Eq.\,(\ref{pipi_fit_be_3dim}) 
an additional term $-2 \qout (2\routlong)^2 \qlong$, where the prefactor accounts for both 
non-diagonal terms, 'out'-'long' and 'long'-'out'.  
We found only marginal differences in the fits which delivered, for all transverse-momentum 
classes, rather small values of $R^2_\mathrm{out\,long}< 1$\,fm$^2$.    
For all results presented here, we restricted 
the pair rapidity to an interval $\vert y - y_\mathrm{cm} \vert<0.35$, 
within which $dN/dy$ does not vary by more than 10\,\%, 
and limited ourselves to the fit function with the Bose-Einstein 
part (Eq.\,(\ref{pipi_fit_be_3dim})) consisting of diagonal elements only and added the small 
deviations to the systematic errors. 
The effect of finite momentum resolutions of the HADES tracking system is studied 
with dedicated simulations. Typical Gaussian resolution values of  
$\sigma_q(\qinv=20\,\mevc) \simeq 2\,\mevc$ are estimated. Incorporating a  
corresponding correction into the fit function 
by convolution of Eq.\,(\ref{pipi_fit_fct_3dim}) with a Gaussian resolution function 
leads to radius shifts of about $\delta R/R \simeq +2$\,\%. 

\begin{figure}
\begin{center}
\includegraphics[width=0.9\linewidth]{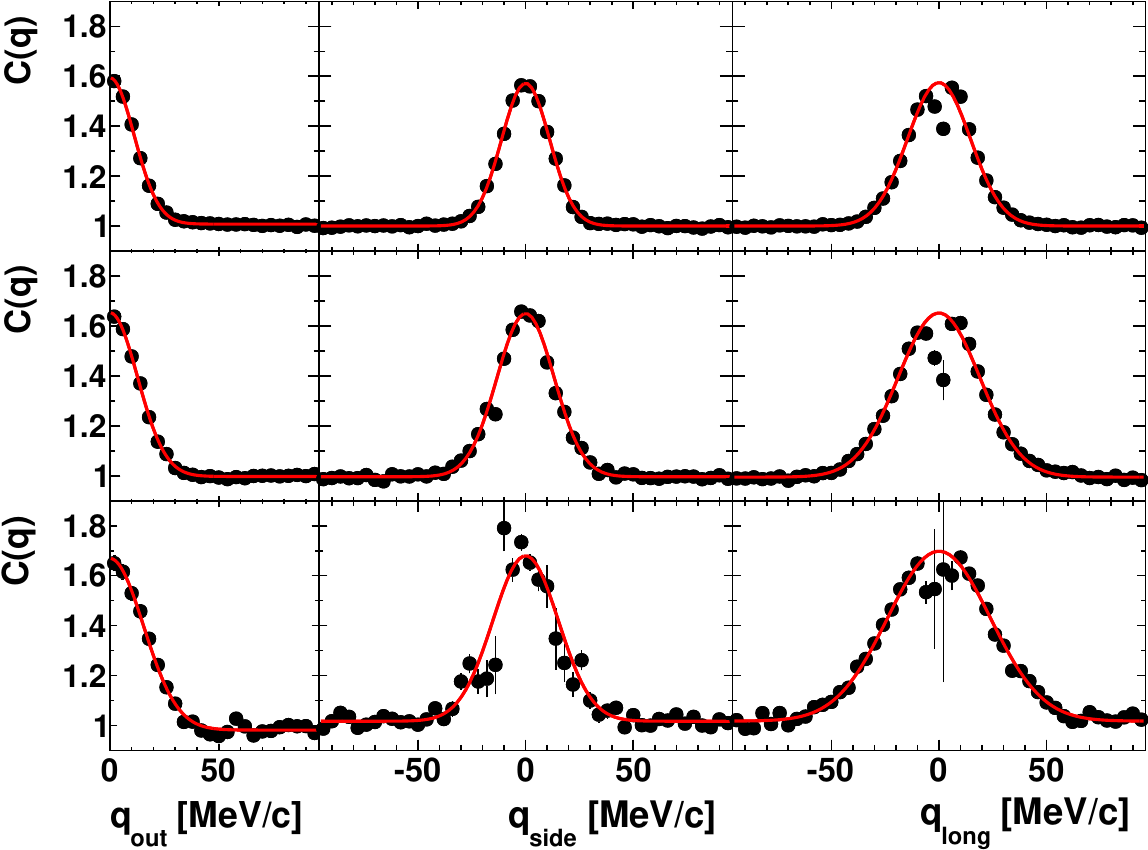}
\end{center}
\caption{Projections of the Coulomb-corrected three-dimensional $\pimin\pimin$ correlation 
function (dots) and of the respective fits (dashed curves) for the $k_\mathrm{t}$ intervals 
of $100-150~\mevc$ (top), $200-250~\mevc$ (middle), and $300-350~\mevc$ (bottom). 
The left, center, and right panels give the 'out', 'side', and 'long' directions, 
respectively. The unplotted $q$ components are integrated over $\pm 12~\mevc$.}
\label{onedim_projections}
\end{figure}

\begin{figure}
\begin{center}
\includegraphics[width=0.9\linewidth]{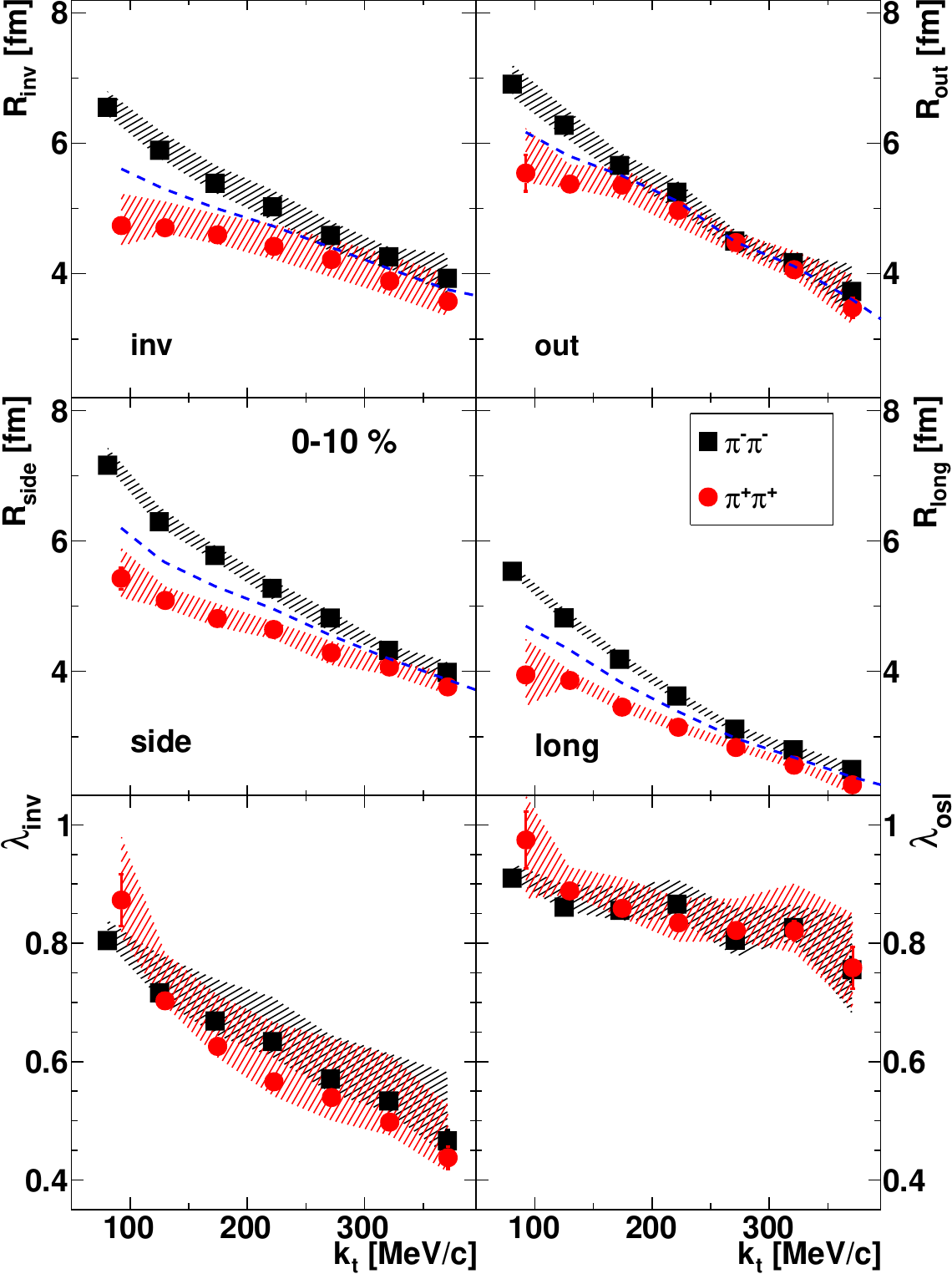}
\end{center}
\caption{Source parameters as function of pair transverse momentum, $k_\mathrm{t}$, 
for central ($0-10\,\%$) \AuAu. 
The upper left, upper right, center left, and center right panels display the invariant, out, 
side, and long radii, respectively. The lower left and lower right panels show the 
corresponding $\lambda$ parameters resulting from the fits to the one- and three-dimensional 
correlation functions, respectively. 
Black squares (red circles) are for pairs of negative 
(positive) pions. Blue dashed lines represent constructed radii of neutral 
pion pairs (see text). Error bars and hatched bands  
represent the statistical and systematic errors, respectively. }
\label{Radien_pi0pi0extract_spline_centr0to10_pt0to900}
\end{figure}

Figure \ref{onedim_projections} shows one-dimensional projections of the Coulomb-corrected 
$\pimin\pimin$ correlation function together with corresponding 
fits with Eq.\,(\ref{pipi_fit_fct_3dim}) for various $k_\mathrm{t}$ intervals. 
(Due to the permutability of particles 1 and 2, one of the $q$ projections 
can be restriced to positive values.) 
The peak due to the Bose-Einstein enhancement becomes evident at low $\vert q \vert$. 
Its width increases with increasing $k_\mathrm{t}$.  The correlation functions for 
$\piplus\piplus$ pairs look similar. 
The main systematic uncertainties of the results presented below arise from the 
slight fluctuations of the fit results when varying the fit ranges ($\sim 0.1 - 0.3$\,fm), 
from the forward-backward differences of the fit results w.r.t. midrapidity within similar 
transverse momentum intervals ($\sim 0.03 - 0.1$\,(0.2)\,fm for $R_{\mathrm{inv}}$, 
$R_{\mathrm{side}}$, $R_{\mathrm{long}}$ ($R_{\mathrm{out}}$)), and from the differences 
when switching on/off the 'out'-'long' component in the fit function ($\sim 0.05 - 0.2$\,fm). 
Finally, all systematic error contributions are added quadratically. 
In Fig.\,\ref{Radien_pi0pi0extract_spline_centr0to10_pt0to900} they are shown as hatched 
bands. 

To separate a potential source radius bias introduced by the Coulomb force the charged pions 
experience in the field of the charged fireball, we follow the ansatz used in 
Ref.\,\cite{Baym_1996}, 
\begin{equation}
E({\bm p}_\mathrm{f}) = E({\bm p}_\mathrm{i}) \pm V_\mathrm{eff}({\bm r}_\mathrm{i}),
\label{eq:baym_1}
\end{equation}
where $E$ is the total energy, 
${\bm p}_\mathrm{i}$ (${\bm p}_\mathrm{f}$) is the initial (final) momentum and 
${\bm r}_\mathrm{i}$ is the inital position of the pion in the Coulomb potential 
$V_{\mathrm{eff}}$ with positive (negative) sign for $\pi^+$ ($\pi^-$). With
\begin{align}
 \frac{R_{\pi^{\pm}\pi^{\pm}}}{R_{\tilde\pi^0\tilde\pi^0}} \approx \frac{q_\mathrm{i}}{q_\mathrm{f}} = \frac{|{\bm p}_\mathrm{i}|}{|{\bm p}_\mathrm{f}|} = \sqrt{ 1 \mp 2 \frac{V_{\mathrm{eff}}}{|{\bm p}_\mathrm{f}|} \sqrt{1 + \frac{m_{\pi}^2}{{\bm p}_\mathrm{f}^2}} + \frac{V^2_{\mathrm{eff}}}{{\bm p}_\mathrm{f}^2}}, 
\label{eq:baym_2}
\end{align}
where $q_\mathrm{i}$ ($q_\mathrm{f})$ is the initial (final) relative momentum, and with  
$V_{\mathrm{eff}} / k_\mathrm{t} \ll 1$, it turns out that the constructed squared source 
radius for pairs of neutral pions (denoted by $\tilde{\pi}^0\tilde{\pi}^0$ in the following  
in contrast to the case where $\pi^{-}\pi^{-}$ and $\pi^{+}\pi^{+}$ data are combined) 
is simply the arithmetic mean of the corresponding quantities of the charged pions,  
\begin{equation}
R_{\tilde\pi^0\tilde\pi^0}^2 = \frac{1}{2} \big(R_{\pi^+\pi^+}^2 + R_{\pi^-\pi^-}^2\big), 
\label{eq:baym_3}
\end{equation}
which is valid for all radius components (even though in the 'out' direction, 
Eq.\,(\ref{eq:baym_2}) looks slightly different).  
Finally, the constructed $\pi^{0}\pi^{0}$ correlation radii 
are derived from cubic spline interpolations of the $k_\mathrm{t}$ 
dependence of both the corresponding experimental $\pi^{-}\pi^{-}$ and $\pi^{+}\pi^{+}$ data. 
This interpolation is necessary because - as result of different detector acceptances - 
the charged pion pairs exhibit slightly different average transverse momenta, 
even though they are measured in identical $k_\mathrm{t}$ intervals. 

Figure\,\ref{Radien_pi0pi0extract_spline_centr0to10_pt0to900} shows the dependence 
on average $k_\mathrm{t}$ (determined for $\qinv<50~\mevc$)
of the one-dimensional (invariant) and three-dimensional source radii for $\pi^-\pi^-$ 
(black squares) and $\pi^+\pi^+$ (red circles) pairs. While for low transverse momentum the 
Coulomb interaction with the fireball leads to an increase (a decrease) of the source size 
derived for negative (positive) pion pairs, at large transverse momentum apparently the 
Coulomb effect fades away. The effect is smallest for $\rout$. Note that the charge 
splitting of the source radii was early predicted by Barz~\cite{Barz96,Barz99} 
who investigated the combined effects of nuclear Coulomb field, radial flow, and opaqueness on 
two-pion correlations for a large collision system such as Au\,+\,Au in the $1 \agev$ energy 
regime. Earlier experimental works at the Bevalac employing a three-body Coulomb correction 
found the effect negligible for their studies of 
smaller systems \cite{Zajc84,Christie92,Christie93}. 
The parameter $\lambda$ derived from the fits with Eq.\,(\ref{pipi_fit_fct_3dim})
appears rather independent of charge sign and decreases only slightly with 
increasing transverse momentum, 
cf. lower right panel of Fig.\,\ref{Radien_pi0pi0extract_spline_centr0to10_pt0to900}. It 
fits well into a preliminary evolution with $\sqrt{s_\mathrm{NN}}$ established 
previously \cite{STAR2015}, except the lowest E895 data point. In contrast, $\lambda$ 
resulting from the fits to the one-dimensional ($\qinv$-dependent) correlation function, 
exhibits a significant decrease with $k_\mathrm{t}$ (cf. lower left panel), 
probably pointing to the fact that the one-dimensional fit function is not adequate. 
Note that deviations from Gaussian source shapes will be studied in a forthcoming paper 
by applying the method of source imaging \cite{BrownDanielewicz97,pipi_imaging_E895}, 
or by using L\'{e}vy source parameterizations \cite{Levy_PHENIX2018}. 

\begin{figure}
\begin{center}
\includegraphics[width=0.9\linewidth]{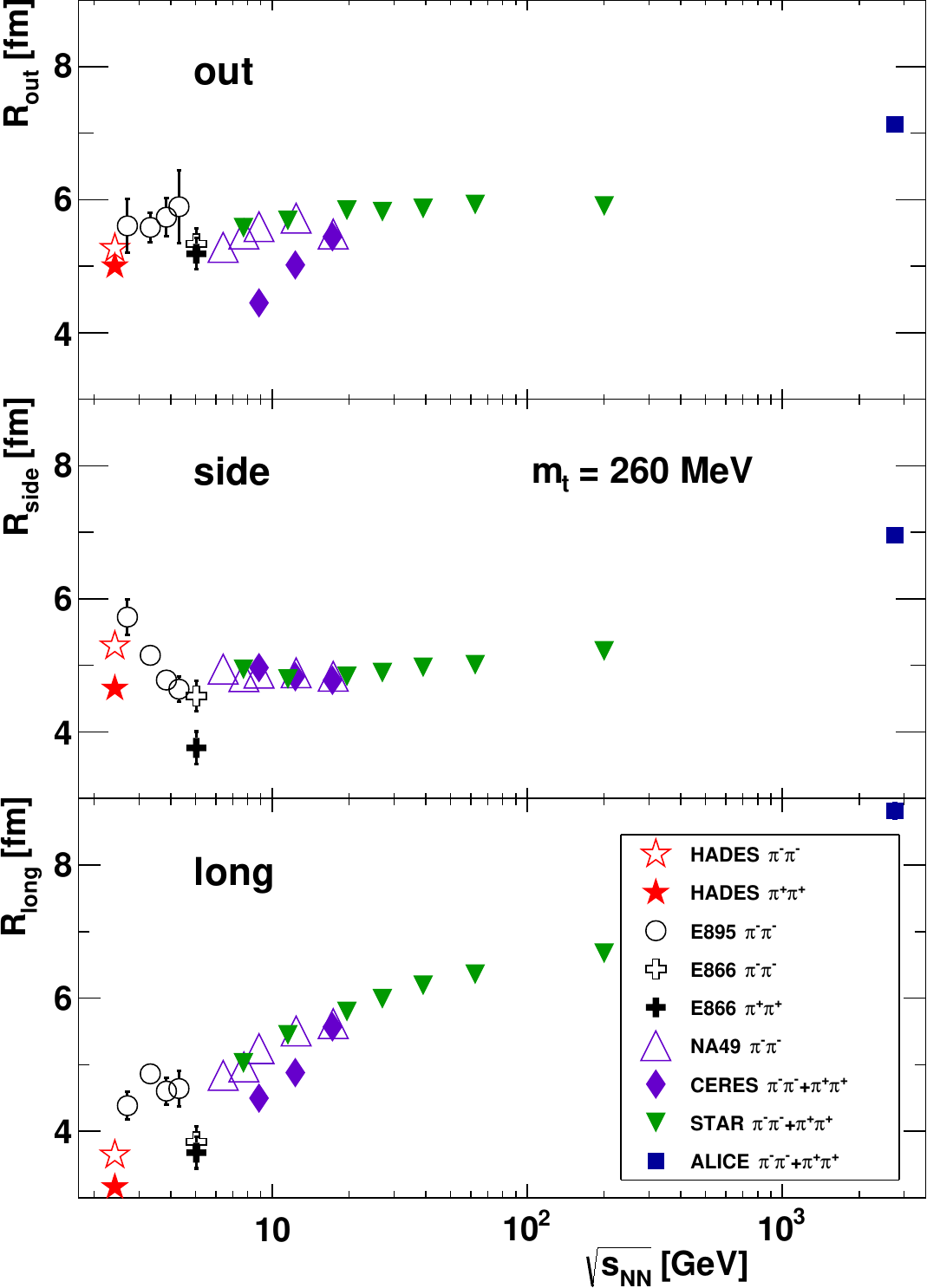}
\end{center}
\caption{Excitation function of the source radii $R_{\mathrm{out}}$ (upper panel), 
$R_{\mathrm{side}}$ (central panel), and 
$R_{\mathrm{long}}$ (lower panel) for pairs of identical pions with transverse mass 
of $m_\mathrm{t}=260\,\mev$ in central collisions of Au\,+\,Au or Pb\,+\,Pb.  
Squares represent data by ALICE at LHC ($\pi^-\pi^-$+$\pi^+\pi^+$) \cite{ALICE2016}, 
full triangles STAR at RHIC ($\pi^-\pi^-$+$\pi^+\pi^+$) \cite{STAR2015}, 
diamonds are for CERES at SPS ($\pi^-\pi^-$+$\pi^+\pi^+$) \cite{CERES_2003}, 
open triangles are for NA49 at SPS ($\pi^-\pi^-$) \cite{na49_2008}, 
open circles are $\pi^-\pi^-$ data by E895 at AGS \cite{e895_2000,Lisa05}, and open 
(full) crosses involve $\pi^-\pi^-$ ($\pi^+\pi^+$) data of E866 at AGS 
\cite{e866_1999}, respectively. The present data of HADES at SIS18 for pairs of 
$\pi^-\pi^-$ ($\pi^+\pi^+$) are given as open (full) stars. Statistical errors are 
displayed as error bars; if not visible, they are smaller than the corresponding symbols. }
\label{Rosl_exctfct_mt260_central_pimpim_pippip}
\end{figure}

The excitation functions of $R_{\mathrm{out}}$, $R_{\mathrm{side}}$, and $R_{\mathrm{long}}$ 
for pion pairs produced in central collisions are displayed 
in Fig.\,\ref{Rosl_exctfct_mt260_central_pimpim_pippip}. All shown radius parameters have been 
obtained by interpolating the existing measured data points to the same transverse mass of 
$m_\mathrm{t}=\sqrt{k_\mathrm{t}^2+m_{\pi}^2}=260\,\mev$ at which data points by STAR at RHIC 
\cite{STAR2015} are available. The statistical errors are properly propagated and 
quadratically added with systematic differences of linear and cubic-spline interpolations. 
Extrapolations were not necessary at this $m_\mathrm{t}$ value.  
Corresponding excitation functions at other 
transverse masses show similar dependencies. Surprisingly, $R_{\mathrm{out}}$ 
and $R_{\mathrm{side}}$ vary hardly more than 40\,\% over three orders of magnitude in 
center-of-mass energy. Only $R_{\mathrm{long}}$ exhibits a systematical 
increase by about a factor of two to three when going in energy from SIS18 via AGS, SPS, RHIC  
to LHC. Note that in the excitation functions of Ref.\,\cite{STAR2015} not all, 
particularly AGS, data points were properly corrected for their $k_\mathrm{t}$ dependence. 
While the HADES $R_{\mathrm{out}}$ and $R_{\mathrm{side}}$ data for negative pions  
completely agree with the lowest E895 data at $2 \agev$,   
$R_{\mathrm{long}}$ deviates from the corresponding E895 data point. 
Both data are, however, in accordance with the overall smooth trend within $2~\sigma$. 
(The low-energy CERES data of $R_{\mathrm{out}}$ and the E866 data point of $R_{\mathrm{long}}$ 
for $\pi^-\pi^-$ pairs appear to be outliers.) 

\begin{figure}
\begin{center}
\includegraphics[width=0.9\linewidth]{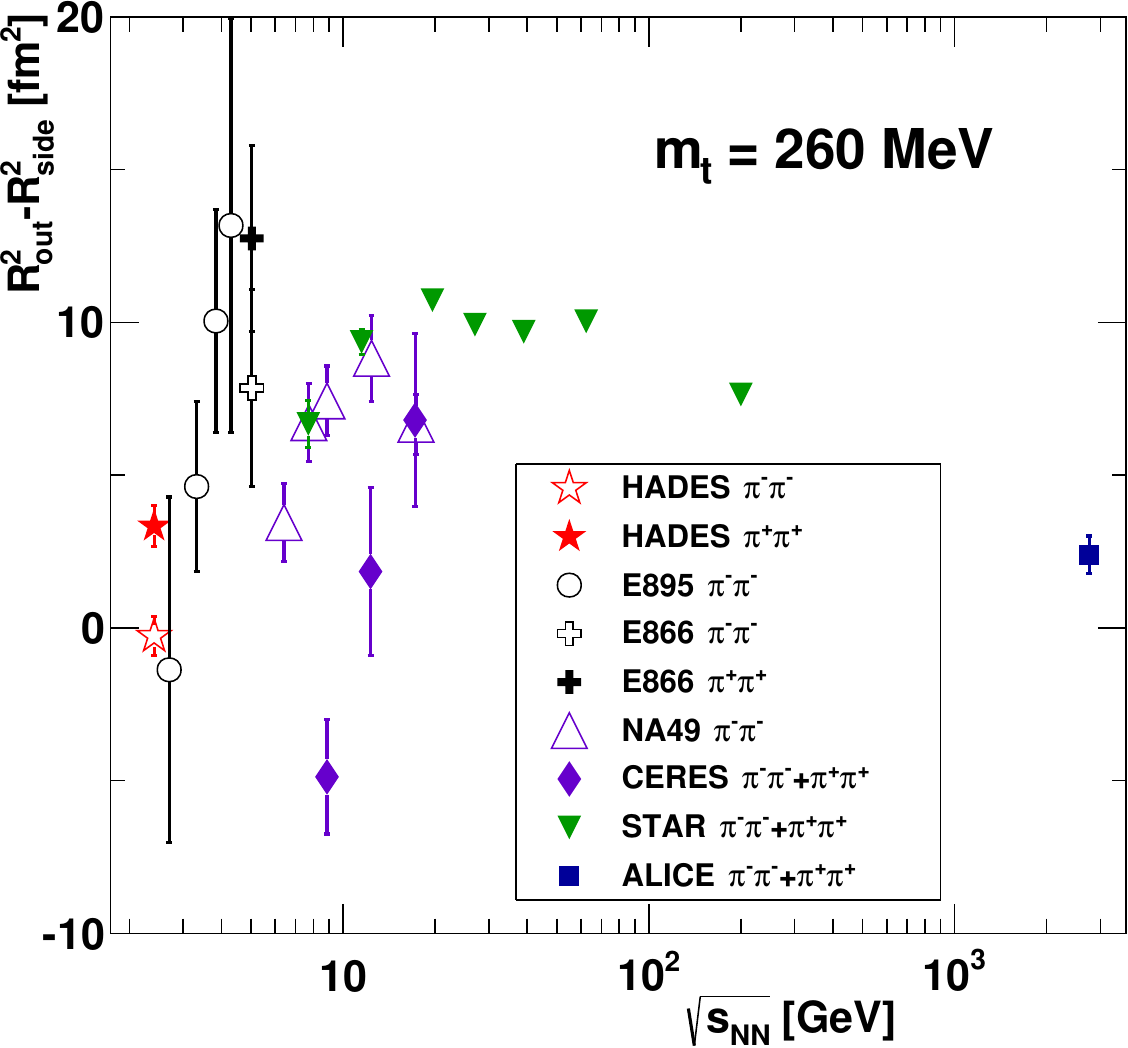}
\end{center}
\caption{Excitation function of $R_{\mathrm{out}}^2-R_{\mathrm{side}}^2$, as calculated from the 
data points shown in Fig.\,\ref{Rosl_exctfct_mt260_central_pimpim_pippip}. }
\label{Rout2_Rside2_mt260_central_pimpim_pippip}
\end{figure}

The combination of $R_{\mathrm{out}}^2$ and $R_{\mathrm{side}}^2$ can be  
related to the emission time duration \cite{Lisa2016}, 
$(\Delta\tau)^2\approx(R_{\mathrm{out}}^2-R_{\mathrm{side}}^2)/\langle\beta_\mathrm{t}^2\rangle$,  
where $\beta_\mathrm{t}$ is the transverse pair velocity.  
The excitation function of $R_{\mathrm{out}}^2-R_{\mathrm{side}}^2$  
is shown in Fig.\,\ref{Rout2_Rside2_mt260_central_pimpim_pippip}.  
Up to now almost all measurements below 10\,GeV are characterized by large errors and scatter 
sizeably. (Here, the outlying low-energy CERES data are solely caused by the 
deviation in $R_{\mathrm{out}}$, cf. top panel of 
Fig.\,\ref{Rosl_exctfct_mt260_central_pimpim_pippip}.)
The new HADES data show that the difference of source parameters in the 
transverse plane almost vanishes at low collision energies. 
With increasing energy, it reaches a maximum at 
$\sqrt{s_{\mathrm{NN}}}\sim 20-30$\,GeV and afterwards decreases towards zero at LHC energies. 
One would conclude that in the $1 \agev$ energy region pions are emitted into 
free space during a short time span of less than one to two fm/$c$. However, also the  
opaqueness of the source affects $R_{\mathrm{out}}^2-R_{\mathrm{side}}^2$ which could cause 
it to become negative, thus compensating the positive contribution from the emission time 
\cite{Barz99}.
 
\begin{figure}
\begin{center}
\includegraphics[width=0.9\linewidth]{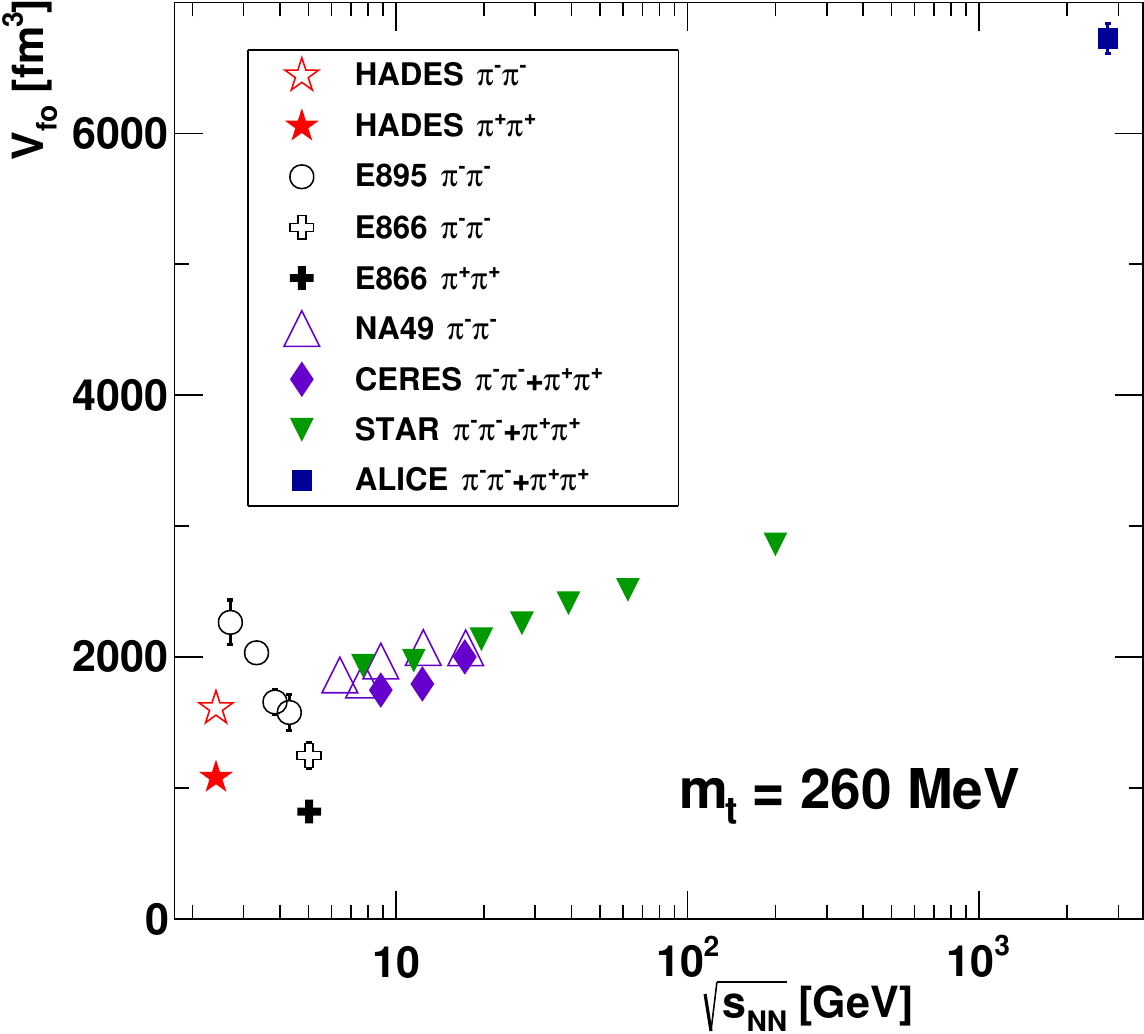}
\end{center}
\caption{Excitation function of the freeze-out volume, 
$V_{\mathrm{fo}}=(2\pi)^{3/2}R^2_{\mathrm{side}}R_{\mathrm{long}}$, as calculated from the 
data points shown in Fig.\,\ref{Rosl_exctfct_mt260_central_pimpim_pippip}. }
\label{Vfreezeout_mt160_central_pimpim_pippip}
\end{figure}

The excitation function of the freeze-out volume, 
$V_\mathrm{fo}=(2\pi)^{3/2}R^2_{\mathrm{side}}R_{\mathrm{long}}$, 
is given in Fig.\,\ref{Vfreezeout_mt160_central_pimpim_pippip}. Note that this definition
of a three-dimensional Gaussian volume does not incorporate $\rout$ since generally this 
length is potentially extended due to a finite value of the aforementioned  emission 
duration. From the above HADES data, we estimate a volume of about 1,300\,fm$^3$ for pairs 
of constructed neutral pions. The volume of homogeneity 
steadily increases with energy, but is merely a factor four larger at LHC. Extrapolating 
$V_\mathrm{fo}$ to $k_\mathrm{t}=0$ yields a value of about 3,900\,fm$^3$. 

The large scatter of data points in 
Fig.\,\ref{Vfreezeout_mt160_central_pimpim_pippip} below $\sqrt{s_{\mathrm{NN}}}= 10$\,GeV is 
intriguing and might indicate a non-trivial energy dependence of the radius parameters in this 
region. However, the simplest interpretation would be to assume instead that the energy 
dependence is smooth. (Note that the difference of the HADES $\pi^-\pi^-$ data 
and the lowest E895 data point at $2\agev$ is primarily caused by the deviation in  
$R_{\mathrm{long}}$.)   
If, however, the variation of the data at low energies, 
most prominently seen in the non-monotonicity of $R_{\mathrm{side}}$ 
(cf. Fig.\,\ref{Rosl_exctfct_mt260_central_pimpim_pippip}), is to be taken seriously, 
new experimental and theoretical efforts are needed to clarify the situation, 
as could be done with the future experiments 
CBM at SIS100/FAIR in Darmstadt \cite{CBM} and MPD at NICA in Dubna \cite{NICA} 
or with the STAR fixed-target program \cite{KMeehan_STAR_2017}. 
Finally, we want to recall that in Figs.\,\ref{Rosl_exctfct_mt260_central_pimpim_pippip},
\ref{Rout2_Rside2_mt260_central_pimpim_pippip}, and \ref{Vfreezeout_mt160_central_pimpim_pippip} 
we display statistical uncertainties only; the systematic ones were not available for all 
experiments. 

In summary, we presented high-statistics $\pi^-\pi^-$ and $\pi^+\pi^+$ HBT data for central 
\AuAu. The three-dimensional Gaussian emission source is studied in dependence on 
transverse momentum and found to follow the trends observed at 
higher collision energies, extending the corresponding excitation functions down to the very 
low part of the energy scale. Substantial differences of the source radii for pairs of 
negative and positive pions are found, especially at low transverse momenta, an effect which 
is not observed at higher collision energies. A clear hierarchy of the three 
Gaussian radii is seen in our data, 
i.e. $R_{\mathrm{long}}<R_{\mathrm{side}}\approx R_{\mathrm{out}}$, independent of transverse 
momentum. Furthermore, a surprisingly small variation of the space-time extent of 
the pion emission source over three orders of magnitude in center-of-mass energy, 
$\sqrt{s_{\mathrm{NN}}}$, is observed. 
Our data indicate that the very smooth trends observed at ultra-relativistic energies 
continue towards very low energies.      
While both $R_{\mathrm{out}}$ and $R_{\mathrm{long}}$ steadily 
decrease with decreasing $\sqrt{s_\mathrm{NN}}$, a weak  
non-monotonic energy dependence of $R_{\mathrm{side}}$ can not be excluded. 

\begin{acknowledgments}
\vskip 10bp
The HADES Collaboration gratefully acknowledges the support by the grants 
SIP JUC Cracow, Cracow (Poland), National Science Center, 2016/23/P/ST2/040 POLONEZ, 
2017/25/N/ST2/00580, 2017/26/M/ST2/00600; TU Darmstadt, Darmstadt (Germany) and 
Goethe-University, Frankfurt (Germany), ExtreMe Matter Institute EMMI at GSI Darmstadt; 
TU M\"unchen, Garching (Germany), MLL M\"unchen, DFG EClust 153, GSI TMLRG1316F, BMBF 05P15WOFCA, 
SFB 1258, DFG FAB898/2-2; NRNU MEPhI Moscow, Moscow (Russia), in framework of Russian 
Academic Excellence Project 02.a03.21.0005, Ministry of Science and Education of the Russian 
Federation 3.3380.2017/4.6; JLU Giessen, Giessen (Germany), BMBF:05P12RGGHM; IPN Orsay, 
Orsay Cedex (France), CNRS/IN2P3; NPI CAS, Rez, Rez (Czech Republic), 
MSMT LM2015049, OP VVV CZ.02.1.01/0.0/0.0/16 013/0001677, LTT17003. 
\end{acknowledgments}

\end{document}